\documentclass[11pt,twoside]{article}

\usepackage{baaa}

\usepackage{graphicx}
\usepackage{subfigure}
\usepackage{amssymb}
\usepackage[spanish,activeacute,english]{babel}
\usepackage[latin1]{inputenc}
\usepackage{verbatim}
\usepackage{amsmath}
\usepackage{amsfonts}
\usepackage{amssymb}
\usepackage[colorlinks=true,dvips]{hyperref}
\usepackage{natbib}

\begin{document}
\myselectspanish
\vskip 1.0cm
\markboth{J. Camperi et al.}%
{Nuevos resultados sobre la cinemática global y
nuclear de NGC 253}

\pagestyle{myheadings}
\vspace*{0.5cm}

\noindent PRESENTACIÓN MURAL

\vskip 0.3cm
\title{Nuevos resultados sobre la cinemática global y
nuclear de NGC 253: movimientos no circulares y emisi\'on en Br\,$\gamma$}


\author{J.A. Camperi$^{1}$, G.I. Gunthardt$^{1}$, R.J. Díaz$^{1,2}$, M. P. Aguero$^{1,2}$, \\G. Gimeno$^{3}$, P. Pessev$^{3}$}

\affil{%
  (1) Observatorio Astron\'omico, Universidad Nacional de C\'ordoba\\ 
  (2) ICATE, CONICET, Argentina\\
  (3) GEMINI Observatory\\
}

\begin{abstract} Continuing with previous research (Camperi et al. 2011), new heliocentric radial velocity distributions are presented for the nearby galaxy NGC 253, obtained from the ionized hydrogen recombination line H$\alpha$. These distributions have been derived from long-slit spectroscopy for various position angles. It is also shown the heliocentric radial velocity distribution corresponding to part of the infrared data (ionized hydrogen recombination line Br$\gamma$) observed with the Phoenix spectrograph of the Gemini South Observatory. Sequential mapping with the long slit using this instrument will enable to study in detail the kinematics of the galaxy's core, which is strongly obscured by dust.
\end{abstract}

\begin{resumen}
  Siguiendo con investigaciones previas (Camperi et al. 2011), se presentan nuevas distribuciones de velocidades radiales heliocéntricas para NGC 253, obtenidas a partir de la línea H$\alpha$ de recombinación del hidrógeno ionizado. Las mismas se consiguieron por medio de la técnica de ranura larga para varios ángulos de posición. Se presenta además una distribución de velocidades radiales heliocéntricas correspondientes a parte de los datos infrarrojos (línea Br$\gamma$ de recombinación de hidrógeno ionizado) observados con el espectrógrafo Phoenix del telescopio Gemini Sur. El mapeo secuencial con la ranura larga permitirá completar el campo de velocidades radiales y la curva de rotación para la región nuclear, no observable en el rango visible pues está fuertemente oscurecida por el polvo.
\end{resumen}

\section{Introducción}
\label{S_intro}

\noindent La gran cantidad de energía radiada por muchos centros galácticos es uno de los puntos claves en el estudio de las galaxias y su evolución. Pese a los avances en este campo, muchos interrogantes siguen presentes hoy en día: ¿La acreción sobre el agujero negro central y la formación estelar violenta son simplemente fenómenos coevolutivos, o se relacionan de alguna forma simbiótica?  ¿Cómo es la física detallada de los mecanismos que disparan la formación estelar violenta extendida en las regiones nucleares?  ¿Cuál es la relación entre estos mecanismos de disparo y la evolución galáctica?

El principal desafío al confrontar estos interrogantes reside en el hecho de que la formación estelar violenta se detecta en estadios avanzados que no proveen suficientes pistas acerca de su origen, puesto que las indicaciones morfológicas del mecanismo de disparo son borroneadas en una escala de tiempo equivalente a unas pocas revoluciones del núcleo galáctico.

En el caso de NGC 253, su cercanía nos permite detectar un estallido de formación estelar violenta en sus inicios, en una etapa en la que aún es posible estudiar en detalle las asimetrías del potencial gravitatorio que han dado origen a la formación estelar violenta, antes de que la misma evolución del fenómeno y la veloz dinámica del centro galáctico las enmascaren (Díaz et al. 2010). A partir de nuestros datos proponemos la existencia de un agujero negro supermasivo fuera del centro cinemático del disco nuclear rico en gas como origen de la perturbación dinámica que desencadena la formación estelar violenta. 

Existen numerosos trabajos observacionales acerca de NGC 253. Por ejemplo, el campo de velocidades del gas ionizado ha sido muestreado por Hlavacek-Larrondo et al. (2011) en gran escala (hasta un radio de 11 minutos de arco, equivalente a 8.5 kpc) con una resolución de 55 segundos de arco.  Sin embargo no hay suficiente información acerca de la distribución espacial de velocidades del gas ionizado y de la componente estelar en el cuerpo central de la galaxia que permita ajustar un modelo detallado de masa. Es fundamental ligar nuestros datos cinemáticos de la región nuclear (que muestrean los 20$''$  de arco centrales a una resolución espacial de 0.2$''$) con la distribución global de masa. Por ello estamos muestreando el cuerpo principal de NGC 253 a través de la técnica de ranura larga secuencial usando observaciones de CASLEO que permiten alcanzar una resolución espacial de 2 a 3$"$ a lo largo de los 3$'$ centrales. Hemos medido ya la velocidad radial del gas ionizado de la región intermedia de la galaxia (ver Camperi et al. 2011), y pretendemos realizar más observaciones hasta completar un campo de velocidad radial completo que ligue la cinemática global con la cinemática circunnuclear y así poder modelar en detalle la coevolución dinámica de la galaxia y el disco circunnuclear incluyendo el supuesto agujero negro descentrado.

\begin{figure}[!ht]
  \centering
  \includegraphics[width=0.4\textwidth]{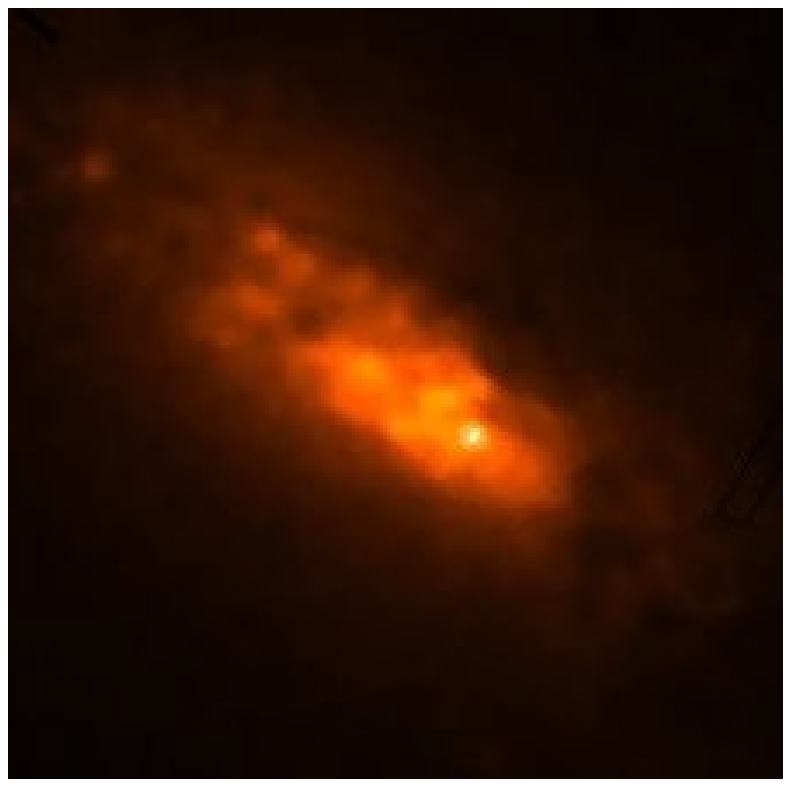}
  \includegraphics[width=0.4\textwidth]{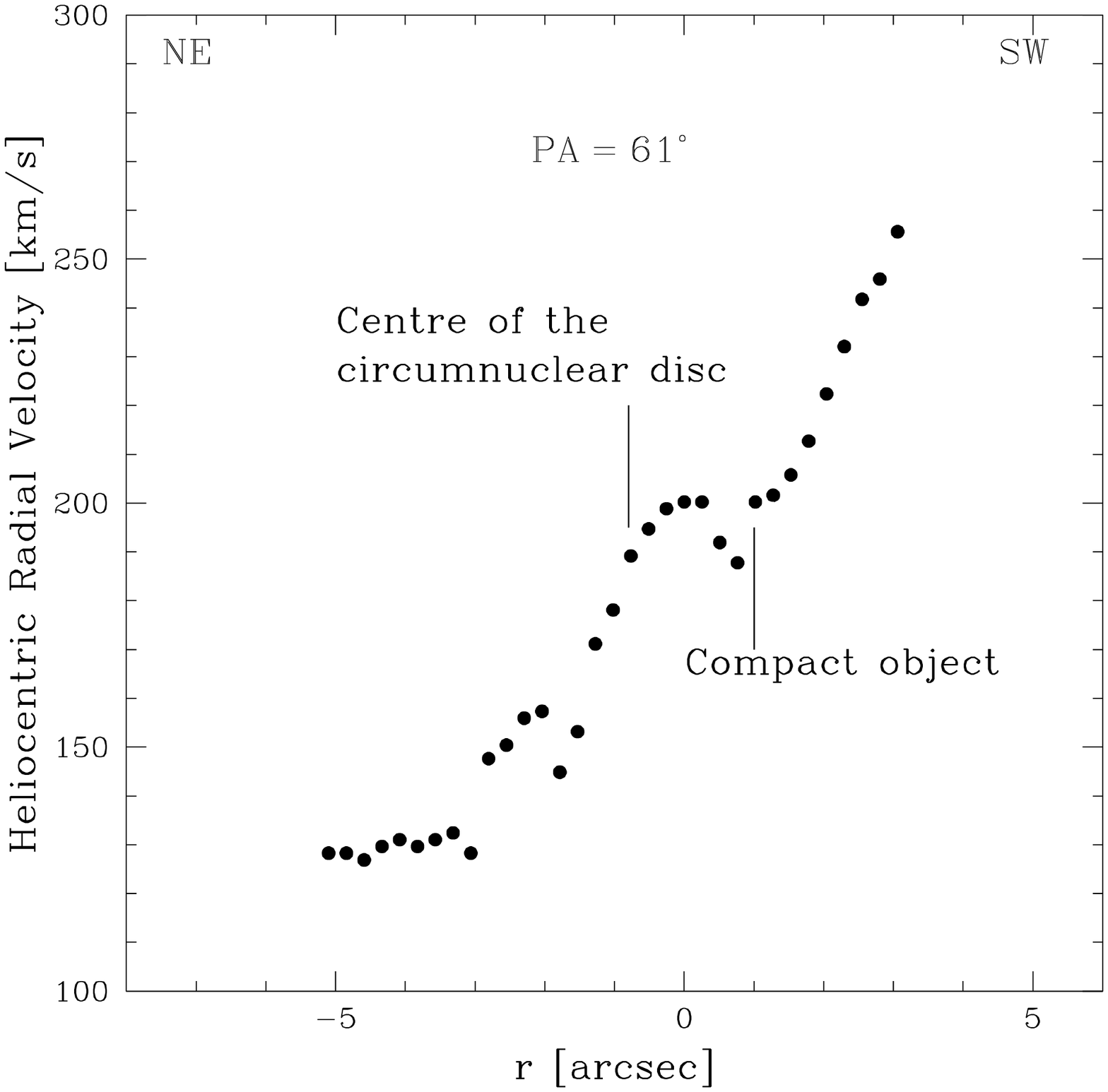}
  \caption{Izquierda: Imagen Phoenix (banda K) del campo estudiado en la región central de NGC 253 ($\approx 14''\times 14''$-N arriba; E a la izquierda). Derecha: distribución de velocidades radiales heliocéntricas de NGC 253 para AP\,=\,61$^{\circ}$ conseguida con el espectrógrafo Phoenix del telescopio Gemini Sur. Las extracciones fueron espaciadas cada 0.25$''$ e integran espacialmente 0.42$''$\,.}
  \label{fig:1}
\end{figure}

\section{Distribuciones de velocidades radiales del gas ionizado en la región intermedia y central de NGC 253}

La noche del 5 de noviembre de 2009 se observó NGC 253 con el espectrógrafo Phoenix del telescopio Gemini Sur de 8 m de Cerro Pachón, Chile. Este espectrógrafo  echelle tiene una resolución espectral de 50000 y un muestreo espacial de 0.085$''$. Se aplicó la técnica de ranura larga sobre la región central de NGC 253 ($\approx 14''\times 14''$). El método observacional consistió en desplazar paralelamente la ranura, cubriendo esta región central, según un ángulo de posición AP\,=\,61$^{\circ}$ (cercano al eje mayor de la galaxia-AP\,=\,50$^{\circ}$). Una de estas posiciones de la ranura, que cruza por sobre dos zonas muy brillantes del disco nuclear fue elegida para obtener una curva de velocidades radiales heliocéntricas correspondiente a AP\,=\,61$^{\circ}$, ajustando perfiles gaussianos a la línea de emisión Br$\gamma$ (Figura 1). 

En octubre de 2009 se realizaron observaciones espectroscópicas  con el espectrógrafo REOSC del telescopio de 2.15 m del CASLEO. Se empleó el modo de ranura larga (sin decker) y una red de difracción de 1200 l/mm. Se obtuvieron  velocidades radiales heliocéntricas para diversos ángulos de posición en la galaxia NGC 253 ajustando perfiles gaussianos a la línea de emisión H$\alpha$. Según Gaspar et al. (2013, este mismo número del Boletín), con la ranura de 4 pixeles (en la zona espectral de 22000 Å, Phoenix) las velocidades de líneas de cielo con S/N $\sim$1 presentan una desviación estándar RMS = 2 km/s. En el caso de REOSC, las velocidades de líneas de cielo con S/N $\sim$5 presentan un RMS de 8 km/s.
En la Figura 2 se muestran las distribuciones de velocidades radiales correspondientes a la dirección AP\,=\,0$^{\circ}$, AP\,=\,27$^{\circ}$, AP\,=\,65$^{\circ}$ y AP\,=\,160$^{\circ}$. Se adicionan así cuatro nuevos AP a los cuatro previamente presentados en Camperi et al. (2011), con lo que ya se dispone de una considerable cobertura propia del campo de velocidades de NGC 253.  Esto nos permitir\'a separar con precisi\'on los flujos radiales de gas con respecto al movimiento circular.
En dos de los AP presentados aquí se revela la presencia de una componente de flujo saliente en la emisión de H$\alpha$, principalmente en dirección sureste, aunque más extendida radialmente que lo reportado previamente por Westmoquette et al. (2011). 

\begin{figure}[!ht]
  \centering
  \includegraphics[width=0.39\textwidth]{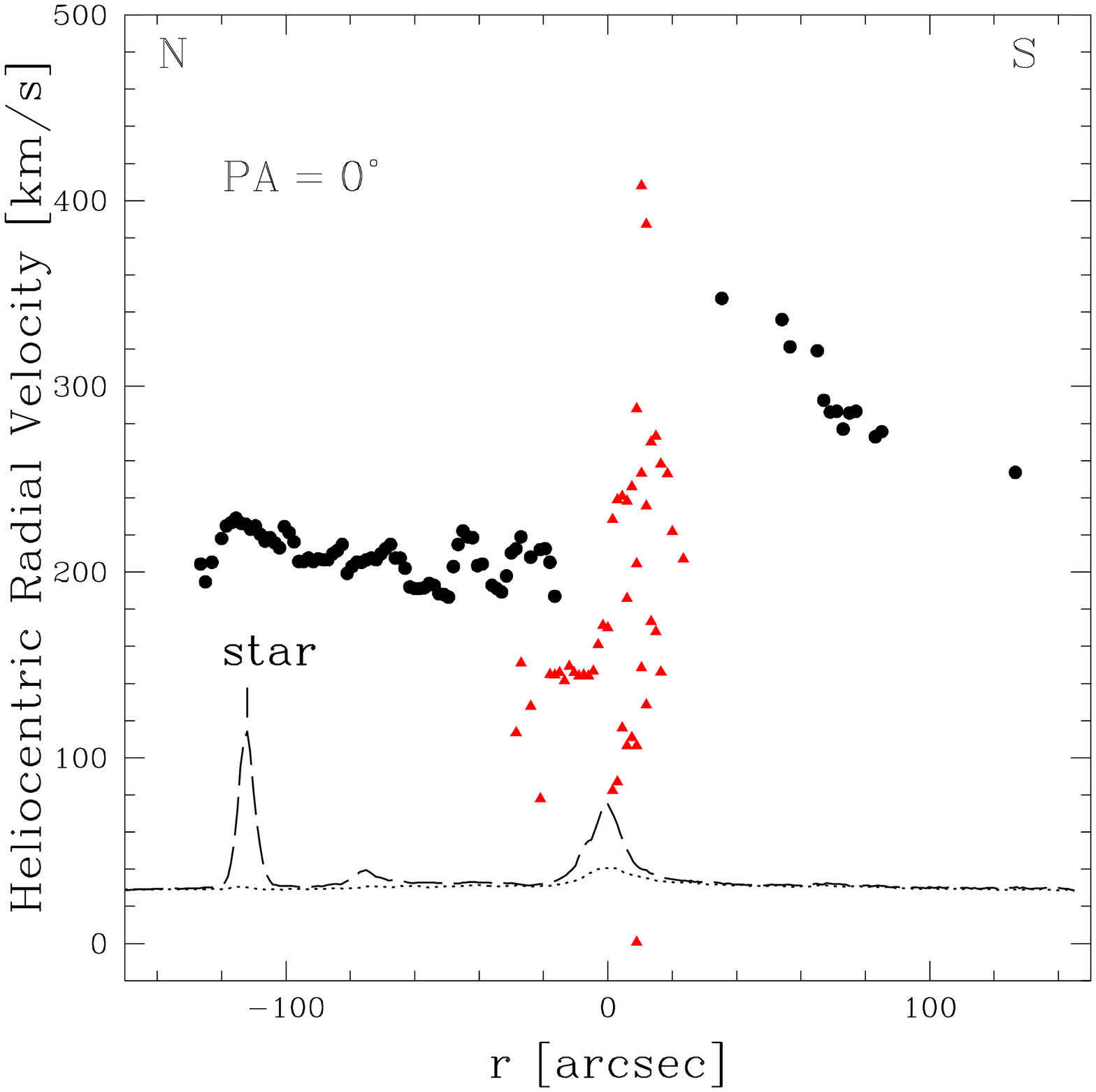}
   \includegraphics[width=0.39\textwidth]{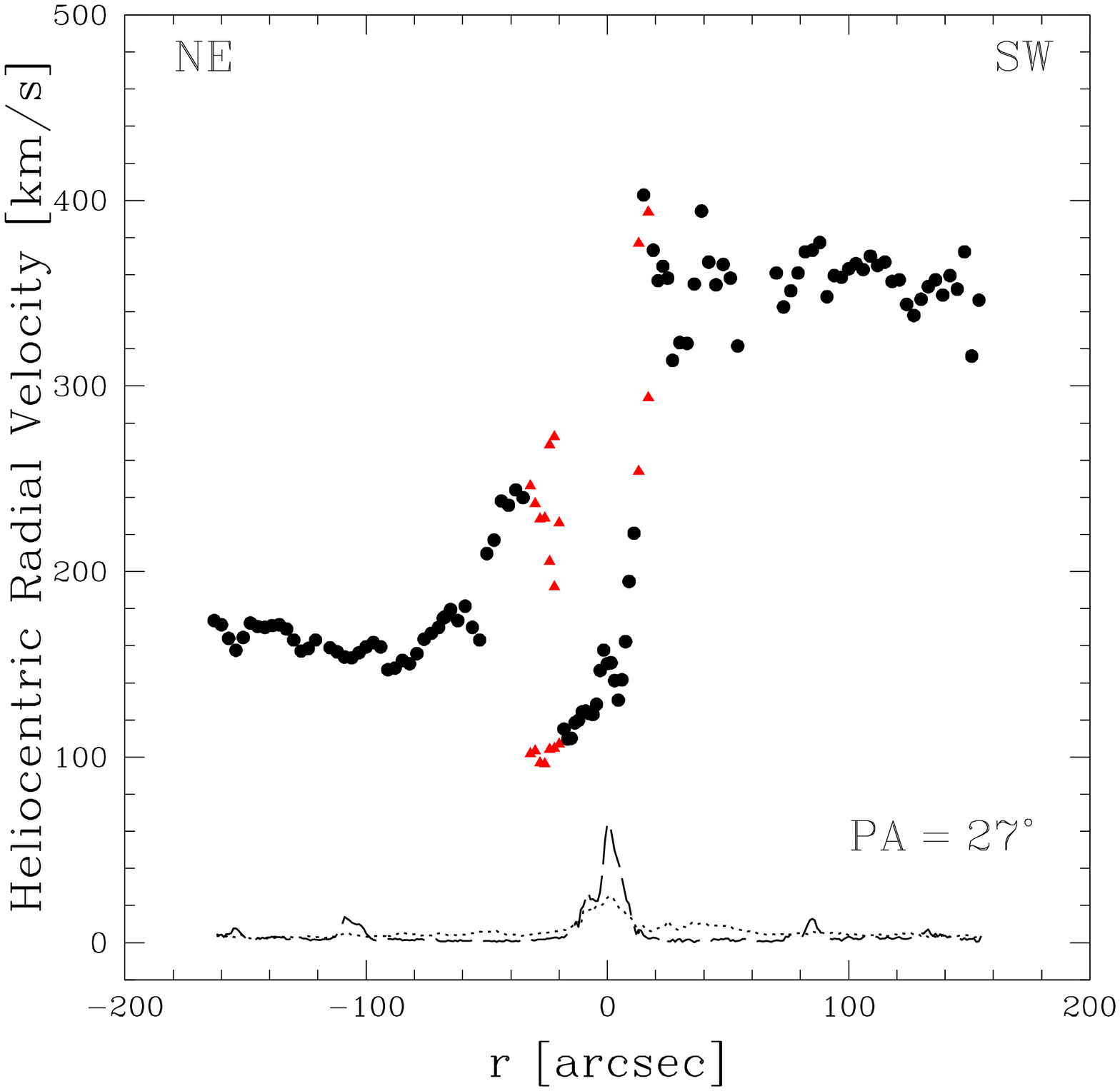}
  \includegraphics[width=0.39\textwidth]{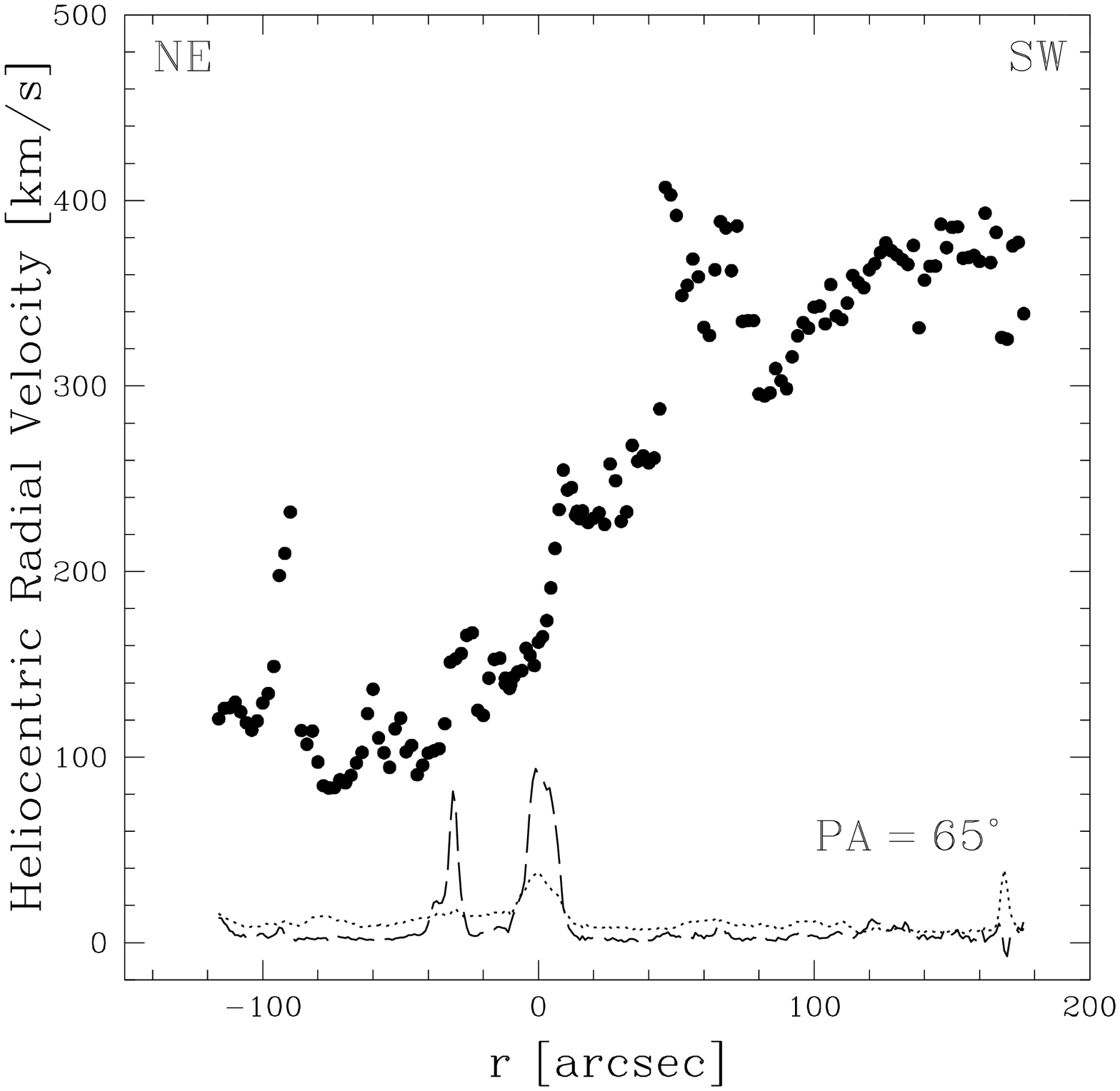}
\includegraphics[width=0.39\textwidth]{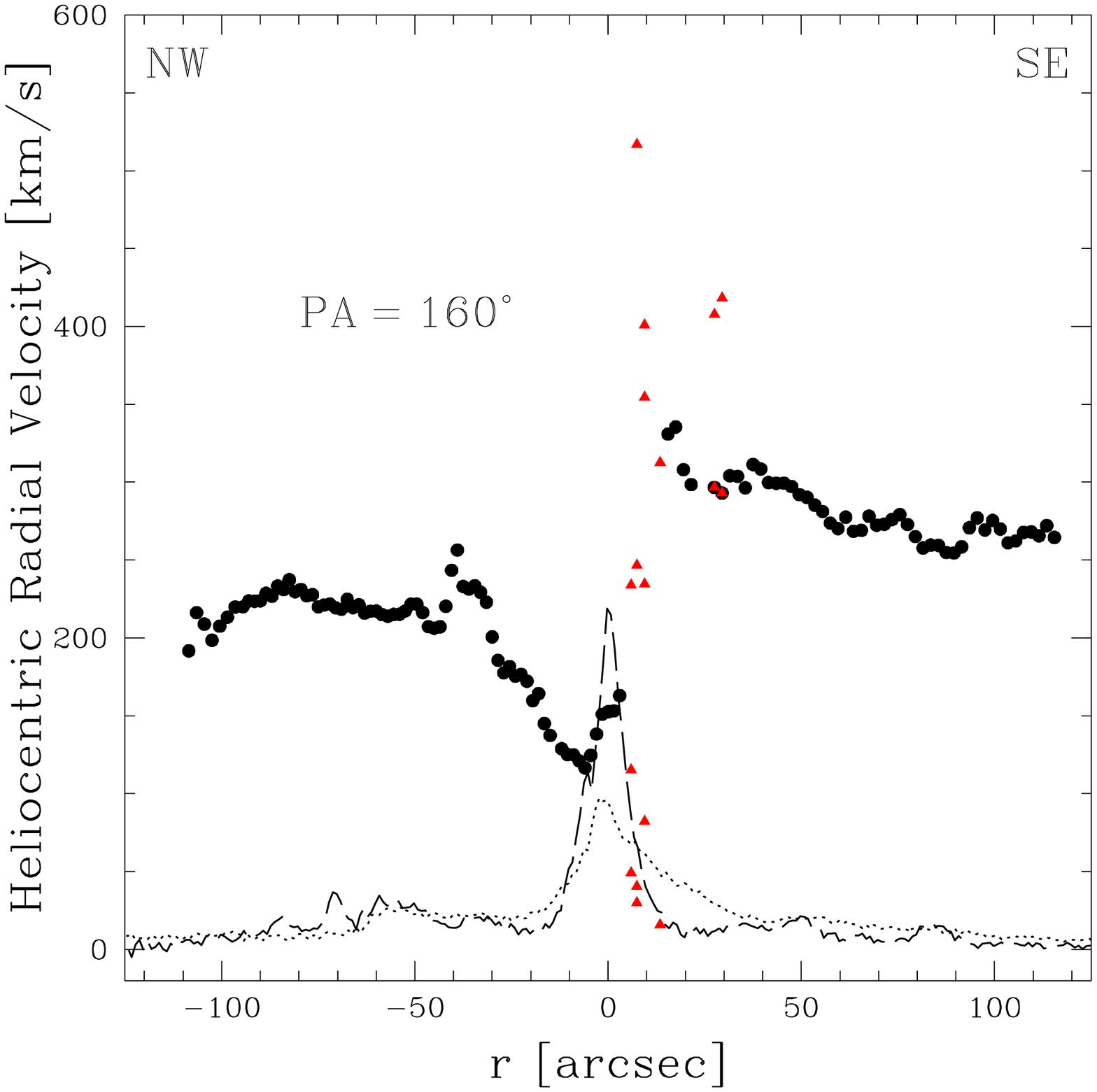}

  \caption{Distribuciones de velocidades radiales helioc\'entricas para nuevos \'angulos de posici\'on de la galaxia NGC 253. En las figuras, los círculos corresponden a la componente más intensa de H$\alpha$ (cuando hay desdoblamiento) y los triángulos a las más débiles. Adicionalmente, se muestran los perfiles del continuo (línea de trazos) y de H$\alpha$ con el continuo sustraído (línea de puntos). Las escalas de estos perfiles son arbitrarias.}
    \label{fig:2}
\end{figure} 

\section{Comentarios}

Los datos de la región central de NGC 253 permitirán constreñir el caso de un agujero negro supermasivo dentro de un disco de formación estelar circunnuclear con un detalle espectroscópico sin precedente. El análisis preliminar de los datos muestra que no hay evidencia de una fuente compacta asociada al centro del disco circunnuclear. El ángulo de posición ya reducido AP\,=\,61$^{\circ}$, correspondiente a las observaciones de Phoenix, cruza el objeto más brillante a unos 4$''$ al SW del centro de simetría global. El objeto más compacto que se puede ajustar al gradiente de velocidades observado está en el rango de 0.8\,x\,10$^{6}$ M$_{\odot}$ a 1.2\,x\,10$^{6}$ M$_{\odot}$, el cual es consistente con el ancho de la componente cinemática de la componente más ancha en la línea de emisión ($\sim$300 km/s) si se asume que corresponde a la rotación dentro del elemento de resolución espacial (6 pc). El objeto más masivo, posiblemente un agujero negro supermasivo, estaría entonces fuera del centro de la galaxia y ubicado en el sector de mayor formación estelar del disco central.

\begin{referencias}
\reference Camperi, J. A., Gunthardt, G. I., Díaz, R. J., A\"guero, M. P., Gimeno, G. \& Pessev, P. 2011, BAAA, 54, 377
\reference Díaz, R. J., Mast, D., Gimeno, G., Dottori, H., Rodrigues, I., Aguero, M. P., Pessev, P. 2010, {\it Light Cores behind Dark Masks}, Galaxies and their Masks, Springer 
\reference Gaspar, G., Díaz R. J., Gunthardt G., Ag\"uero, M. P., Camperi, J. A., Gimeno G. 2013, BAAA, 55, en prensa.
\reference Hlavacek-Larrondo, J., Carignan, C., Daigle, O., de Denus-Baillargeon, M.-M., Marcelin, M., Epinat, B., Hernandez, O. 2011, MNRAS 411, 71
\reference Phoenix Spectrometer:  http://www.gemini.edu/?q=node/10239
\reference Westmoquette, M.S., Smith L. J. \& Gallagher III, J. S. 2011, MNRAS, 414, 3719
\end{referencias}

\end{document}